# Smectic-*A* and Smectic-*C* Phases and Phase Transitions in $\bar{8}S5$ Liquid Crystal-Aerosil Gels

B. Freelon[1], M. Ramazanoglu[2], P.J. Chung[3], P. Valdivia[4], C. W. Garland[5], and R. J. Birgeneau[1,4,6]

[1]*Department of Physics, University of California, Berkeley, CA, 94720*

[2]*Department of Physics, McMaster University, Hamilton Ontario, ONL8S 4M1 Canada*

[3]*Department of Physics, University of California, Santa Barbara, Santa Barbara, CA 93106*

[4]*Department of Materials Science and Engineering, University of California, Berkeley, Berkeley, CA 94720*

[5]*School of Science, Massachusetts Institute of Technology, Cambridge, MA 02139*

[6]*Materials Science Division, Lawrence Berkeley National Laboratory, Berkeley, CA 94720*

High-resolution X-ray scattering studies of the nonpolar thermotropic liquid crystal 4-*n*-pentylphenylthiol-4'-*n*-octyloxybenzoate ( $\bar{8}S5$ ) in aerosil gel nano-networks reveal that the aerosil-induced disorder significantly alters both the nematic to Smectic-*A* and Smectic-*A* to Smectic-*C* phase transitions. The limiting $\bar{8}S5$ Smectic-*A* correlation length follows a power-law dependence on the aerosil density in quantitative agreement with the limiting lengths measured previously in other Smectic-*A* liquid crystal gels. The Smectic-*A* to Smectic-*C* liquid crystalline phase transition is altered fundamentally by the presence of the aerosil gel. The onset of the Smectic-*C* phase remains relatively sharp but there is an extended coexistence region where Smectic-*A* and Smectic-*C* domains can exist.

## I. INTRODUCTION

Quenched random disorder (QRD) has an ubiquitous influence on the physical properties of materials as varied as ferroelectrics, magnets, and superconductors. In particular, QRD may modify in a fundamental way the phase transition behavior in condensed matter systems including both solids and soft matter materials. Among soft matter systems, thermotropic liquid crystals (LCs) have long been used as model systems to explore the nature of phase transitions because they easily undergo phase changes due to a variation of temperature, and both the order parameters and fluctuations are readily accessible experimentally. QRD can be controllably introduced into LCs by techniques [1,2] that confine the molecules inside porous random aerosil gel networks. In such cases, thermotropic liquid crystals which are mixed with aerosils to create a low density gel state are subsequently driven through molecular orientational and structural ordering changes in order to investigate the phase transition behavior.

The inclusion of aerosil nanoparticles in LCs by appropriate chemical procedures [2] results in the formation of a low volume-fraction silica ($SiO_2$) gel network. The random nanoscale gel network's contact interactions with the LCs act as random local pinning fields for both the orientational and positional order. The random field varies over length scales much smaller than those of the LC ordered phases. Thus, LC + aerosil gels provide an excellent experimental window through which to observe the manner in which QRD affects phase transitions. A key feature of LC + aerosil systems is that the degree of random field disorder is proportional [3] to the reduced aerosil mass density, in grams of $SiO_2$ per $cm^3$ of LC, $\rho_S$. By changing $\rho_S$, the random field disorder varies continuously from the dilute ($\rho_S < 0.01$ g/cm$^3$) to the soft ($0.01 < \rho_S < 0.1$ g/cm$^3$) and, at large $\rho_S$, to the stiff ($\rho_S > 0.1$ g/cm$^3$) gel disorder limits.

The influence of random gels on LCs has been primarily studied for the nematic (*N*)-to-Smectic-*A* (Sm*A*) transition. The most salient result of these studies is that the quasi-long-range Sm*A* ordering of the LCs is destroyed by an aerosil gel. [4,5] Even minimal amounts of quenched disorder, that is, very small $\rho_S$, adversely affect the formation of smectic order. [6] Thus, no true Sm*A* phase exists and the smectic correlation length remains finite for all temperatures and values of $\rho_S$ that have been studied. [5] The *N*-Sm*A* phase transition is in the *XY* universality class; however, some systems seem to show a 3D-tricritical to 3D-*XY* crossover behavior for the *N*-Sm*A* transition with increasing disorder strength. [7] In certain liquid crystal systems, as the temperature is lowered below the Sm*A* phase a transition to a tilted or smectic-*C* (Sm*C*) molecular phases occurs; in a previous study, the smectic correlation lengths have been found to either saturate or diminish on cooling into the Sm*C* phase. [8]

Recently, the role of QRD on tilted smectics (Sm*C*) has been further investigated. Technological inquiries have focused on potentially practical magnetic and optical properties of LCs in random environments, while experimental [8, 9,10,11] and theoretical [12] investigations of the Sm*A*-Sm*C* phase transitions within confined geometries have been conducted. Both Sm*A* and Sm*C* phases are one-dimensional mass density waves in three dimensional fluids that possess molecular orientational ordering within the layers formed by the remaining two dimensions. [13, 14] In the Sm*C* (Sm*A*) phase the orientational axis of the molecules is tilted (parallel) relative to the smectic layer normal vector, $\mathbf{n}_0$. The formation of the Sm*C* phase can be characterized by a two-component order parameter which consists of the magnitude of the tilt and its azimuthal angle. [15] The tilt angle $\Phi$, *c.f.* Fig. 1(a), is the primary order parameter of the Sm*C* phase. Detailed high-resolution calorimetry work [16] on 4-*n*-pentylphenylthiol-4'-*n*-octyloxybenzoate ($\overline{8}S5$) in aerosil gels shows that both the *N*-Sm*A* and Sm*A*-Sm*C* phase transitions are dramatically affected by the presence of the aerosil gel. In contrast to the calorimetry work, Clegg *et. al.* [8] presented high-resolution X-ray diffraction data that suggested that both the Sm*A* and Sm*C* phases and phase transitions were largely unaffected by the random fields originating from the aerosil gel network. This finding is also in disagreement with the results of subsequent lower resolution X-ray diffraction studies of doping-induced trends observed for $\overline{8}S5$ [10] and with the phenomenology deduced from studies of a number of other LC-aerosil gel systems. Below, we address the discrepancies regarding the Sm$A$ and Sm$C$ phases and phase transitions in $\overline{8}S5$ + aerosil gel systems.

## II. EXPERIMENTAL RESULTS

$\overline{8}S5$ is a nonpolar liquid crystal whose sequence of phase transitions is shown in Fig. 1(b). Above the isotropic ($I$) to nematic transition temperature ($T_{IN}$ = 359.3 K) the molecules are randomly ordered both positionally and orientationally. A transition from the orientationally ordered, but translationally invariant $N$ phase to the Sm$A$ phase occurs at $T_{NA}$ = 336.6 K. Lowering the temperature brings about the Sm$A$-to-Sm$C$ transition at $T_{AC}$ = 329.3K. Bulk $\overline{8}S5$ has two continuous $XY$ phase transitions. One is the fluctuation-dominated $N$-Sm$A$ transition and the other is the Landau mean-field (MF) Sm$A$-Sm$C$ transition associated with changes in the tilt angle. Low-resolution X-ray diffraction performed on $S$-(+)-[4-2'-methyl butyl] phenyl 4'-$n$-octylbiphenyl-4-carboxylate (CE8) + aerosil [9,10] and $p$-($n$-decyloxybenzylidene)-$p$-amino-(2-methyl-butyl) cinnamate (DOBAMBC) + aerosil [10] revealed that the presence of the aerosil gel resulted in the smearing of the Sm$A$-Sm$C$* transition for those polar and chiral LCs. These studies also suggested the formation of pretransitional Sm$C$ order well inside of the Sm$A$ regime for such LC systems.

In this paper we present a systematic study of the effect of varying the hydrophilic aerosil gel density $\rho_S$ on the $N$-Sm$A$ and the Sm$A$-Sm$C$ phase transitions of $\overline{8}S5$. The Sm$A$ correlation length is found to depend on $\rho_S$ in a manner that is quantitatively consistent with the behavior observed in other LC + aerosil systems. In agreement with several other recent studies of the influence of the aerosil gel on the Sm$A$-Sm$C$ transition, we observe that the aerosil gel dramatically alters the onset behavior of $\overline{8}S5$ Sm$C$ ordering. In addition, we find that quenched random disorder not only modifies the Sm*C* order, but also leads to the formation of Sm*C* domains which can coexist with Sm*A* domains over an extended temperature range, where presumably the nature of the domains is determined by the local gel geometry.

We should note that for isotropic liquid crystal-aerosil gel systems, the descriptors *N*, Sm*A* and Sm*C* used in this paper are not technically correct. Because of the quenched random fields exerted by the gel on both the orientational and positional order, none of these phases has long-range order; that is, they are all actually in the *I* phase. Nevertheless, as this and previous studies show, there is pronounced short range orientational and positional order. It is this short range order that the labels *N*, Sm*A* and Sm*C* refer to in this work.

We studied $\overline{8}S5$ which was purchased from the Military University of Technology in Warsaw. [17] The molecular shape and structure of $\overline{8}S5$ are shown in Fig. 1(c,d). The transition temperatures of this LC matched those previously published for $\overline{8}S5$. Hydrophilic silica ($SiO_2$) 7-nm diameter nanospheres [3], Aerosil 300, were obtained from Degussa Inc. [17] The degree of disorder, which is proportional [3] to $\rho_S$, was varied according to preparation techniques outlined in Refs. 16 and 18 to obtain samples with the following mass densities: 0.025, 0.039, 0.060, 0.092, 0.140 and 0.264 g $SiO_2/cm^3$LC. High-resolution X-ray scattering was performed at beamline 2-1 of the Stanford Synchrotron Radiation Laboratory. The 10 keV photon beamline contained double-bounce monochromators, Si(111) analyzer crystals and automated filters which allowed direct beam measurements of the resolution function. We performed transmission powder diffraction scans in which the scattering vector magnitude $q$ was varied from 0.07 to 0.4 Å$^{-1}$. A high longitudinal momentum resolution, $\Delta q \sim 7 \times 10^{-4}$ Å$^{-1}$, was used to ensure that the instrumental resolution was much narrower than any of the measured X-ray scattering peaks. Beam damage of the samples was avoided by the judicious selection of X-ray filters and by limiting the total number of temperatures at which scattering data were collected.

X-ray scattering is uniquely suited to probe the structural ordering of liquid crystals embedded in random aerosil gel networks by providing direct access to the onset and development of quasi-Bragg peaks that result from smectic layering. In order to assess accurately the amplitude and correlation length of the Sm$A$ ordering, we conducted a quantitative analysis of the X-ray scattering lineshapes as a function of temperature and aerosil density. The reproducibility of the X-ray scattering profiles was tested and confirmed for selected sample densities $\rho_S$. Our analysis was guided by the theoretical framework for liquid crystals in quenched random field environments. [6] Because random confinement imposes both tilt and positional disordering fields that couple linearly to the nematic director and smectic order parameter, respectively, detailed information regarding the Sm$A$ correlation length can be gained by constructing an appropriate scattering structure factor, $S(\boldsymbol{q})$. We use what is, in essence, a microcrystalline model for the structure factor by starting with a two-part structure factor, $S(\boldsymbol{q})$, which consists of a Lorentzian and a Lorentzian squared. [19,20,21,22,23] The structure factor is written as the sum of a thermal (high-temperature) part and a static (low-temperature) part $S(\boldsymbol{q}) = S^{thermal}(\boldsymbol{q}) + S^{static}(\boldsymbol{q})$ such that

$$S^{thermal}(\boldsymbol{q}) = \frac{\sigma_1}{1+\left(q_\parallel - q_0\right)^2 \xi_\parallel^2 + q_\perp^2 \xi_\perp^2 + c q_\perp^4 \xi_\perp^4} \qquad (1)$$

$$S^{static}(\boldsymbol{q}) = \frac{a_2\left(\tilde{\xi}_{\parallel 2} \tilde{\xi}_{\perp 2}^2\right)}{\left[1+\left(q_\parallel - q_0\right)^2 \tilde{\xi}_{\parallel 2}^2 + q_\perp^2 \tilde{\xi}_{\perp 2}^2\right]^2} \qquad (2)$$

where $\sigma_1$ is the thermal amplitude and is proportional to the connected susceptibility of the material.  The position of the scattering peak $q_0$ is inversely proportional to the smectic layer spacing periodicity $d$.  $q_\parallel$ ($q_\perp$) is the scattering momentum component that is parallel (perpendicular) to the smectic layer normal.  The thermal correlation lengths $\xi_\parallel$ and $\xi_\perp$ are parallel and perpendicular to the direction of the smectic mass density wave vector, **k**.  $S(\boldsymbol{q})$ is a combination of a Lorentzian Eq. (1) and a squared-Lorentzian Eq. (2) with an additional fourth order term $\sim q_\perp^4$ in the former that originates from splay-mode [24] director fluctuations.  The coefficient $c$ was found to have a value of 0.25 in Ref. 5, and that value has been used in the present fitting procedure.  Two important assumptions are placed upon Eqs. (1) and (2).  First, the *static* longitudinal correlation length $\xi_{\parallel 2}$ is assumed to be different [6] from the *thermal* correlation length $\xi_\parallel$.  Second, the static correlation lengths used in Eq. (2) have been fitted to values which are assumed to be constant for all temperatures, which is indicated by the tilde.  Therefore the static longitudinal correlation length $\tilde{\xi}_{\parallel 2}$ is distinct from the thermal correlation length $\xi_\parallel$.  Because the liquid crystal is embedded in a gel network, the macroscopic system is isotropic as noted above; accordingly the microcrystalline structure factor, Eq. (1) + Eq. (2), must be spherically averaged before comparing it to the measured longitudinal X-ray scans.

In the results presented here, we have closely followed the approach that was given in Refs. 3, 5, 25, 26 .  As for the scattering contribution of the aerosil network, small-angle X-ray scattering has shown [2] that the aerosil gel gives rise to a Porod background; therefore, the aerosil gel background function, $B(q)$, has the form $B(q) = b_P/q^4 + b_C$, where $b_P$ and $b_C$ are constants.  We reduce the number of adjustable parameters by assuming that the ratio of the transverse to longitudinal lengths is identical to that in the bulk.  Since we measure only longitudinal profiles in a spherically symmetric scattering cross-section this assumption has only a minimal effect on the data analysis. The fits, therefore, involve as adjustable fitting parameters $\sigma_1(T), \xi_\|(T), a_2(T), q_0(T), \tilde{\tilde{\xi}}_{\|2}$ in addition to the temperature-independent background due to the aerosil gel network. X-ray scattering lineshapes at each temperature were fitted to the convolution of $S(q)$, which again is the spherical average of Eq. (1) + Eq. (2), and the instrumental resolution function. Subsequently, a global fit to all scans was performed that simultaneously adjusted, at all temperatures, the above parameters in order to achieve a better overall consistency of the fit parameters. In this manner, the static Sm$A$ correlation lengths, $\tilde{\tilde{\xi}}_{\|2}$, were extracted as global fit parameters for each sample of aerosil density $\rho_S$. A representative set of scans with the best fit theoretical line-shapes is shown in Fig. 2. A log-log plot of the $\rho_S$ dependence of the static (Sm$A$) parallel correlation length of $\overline{8}S5$ + aerosil so-obtained is given in Fig. 3. These results differ significantly from those of Clegg *et. al*. [8] who report much larger correlation lengths. We will comment on this discrepancy later in this paper.

Also shown in Fig. 3 are data from several other liquid crystal-aerosil systems which have been studied previously. It is evident that there is good absolute agreement between the lengths measured in these different systems. The data from all of the LC + aerosil samples were fitted together, by least-squares minimization, to obtain an overall characterization of the Sm$A$ correlation length versus $\rho_S$. The obtained fit of the combined data yields a power-law dependence, $\xi_{\|2} \sim \rho_S^{-y}$, where $y \sim 1.14 \pm 0.14$. Thus, the Sm$A$ correlation length is roughly inversely proportion to the disorder strength. [27] The collapse of the data from various LC + aerosil systems provides further evidence that the effect of aerosils on the $N$-Sm$A$ phase transition is similar for polar and non-polar as well as bent-core and rod-like LC molecules.

## III. DETAILS OF THE SmA-SmC TRANSITION

Information about the tilted phase was obtained by cooling the $\overline{8}S5$ + aerosil gel samples from the SmA to the SmC temperature regime. Fig. 4(a) shows a longitudinal X-ray diffraction scan containing a single SmA scattering peak that results from the SmA ordering deep within that phase region for a sample with $\rho_S = 0.025$ g/cm$^3$. Upon lowering the temperature, a second quasi-Bragg peak emerges as SmC order develops, as shown in panel (b) of Fig. 4. The two peaks are located at distinct wave vectors and signify a two-phase region corresponding to coexisting SmA and SmC phases in the sample. Due to the larger layer spacing in the SmA phase, the associated SmA scattering peaks are located at lower $q$ positions than the SmC peaks. Panel 4(c) displays a lower temperature profile in which the coexistence peaks have switched intensity magnitudes because of SmC region growth within the sample. A lone SmC phase peak is shown (Panel 4d) for temperatures below the coexistence region. The lineshape of the SmC quasi-Bragg peaks is qualitatively different from that of the SmA phase as can be seen from a comparison of the profiles in panels (a) and (d) of Fig. 4. The coexistence temperature range was estimated by determining the temperatures over which SmA and SmC coexistence peaks were clearly visible in the intensity profiles of each sample. These estimates therefore represent lower bounds. A finite-range two-phase coexistence of smectic phases has been reported [28,29,30] previously for first-order transitions. However, the SmA + SmC behavior presented here indicates a different behavior in the aerosil gel materials. We now discuss the nature of the coexistence phases by considering the SmC order parameter behavior of the various doped samples.

A simple molecular rod model [13] of the SmC order parameter $\Phi$ (molecular tilt angle) gives $\Phi = \cos^{-1}\left(d_{Sm-C}/d_{Sm-A0}\right)$ where $d_{Sm-A0}$ is the layer spacing of the SmA phase and $d_{Sm-C}$ is the layer spacing in the SmC phase.  $\Phi$ is measured as the angle subtended by the nematic director and the smectic layer normal $\mathbf{n}_0$.  Both lengths are extracted from the data by simply assuming Bragg's law $\lambda = d_{SmA(SmC)}\sin 2\theta_{A(C)}$, where $2\theta_{A(C)}$ are the angular positions of the SmA(C) diffraction peaks.  The primary SmC order parameters versus temperature so-obtained are shown for several $\overline{8}S5$ + aerosil samples ($\rho_S = 0, 0.025, 0.039, 0.060, 0.092$ and $0.264$) in Fig. 5.

In the SmA + SmC disorder-induced coexistence region, several overlapping SmC peaks were observed for high $\rho_S$ samples. [31]  This is presumably due to there being a distribution of LC domains in the gel "pore" structure with different local $T_{AC}$ transition temperatures and distinct tilt angles at a given temperature $T$ near $T_{AC}$.  When this multi-peak scattering is observed, the tilt angles shown in Fig. 5 are based on the prominent features at the largest scattering angle (*i.e.*, the largest tilt angle).  The range of the two-phase coexistence region is indicated by the vertical lines in Fig. 5.  It should be noted that the upper bounds for the coexistence regions are quite close to $T_{AC}^o$, the SmA-SmC transition temperature for pure $\overline{8}S5$, except for the sample with $\rho_S = 0.039$.  Indeed, the entire temperature scale for that sample seems low by ~ 6K compared to samples with $\rho_S = 0.025$ and $0.060$, although we have no explanation for such an anomaly.

The SmA-SmC phase transition in pure $\overline{8}S5$ is expected [32] to be a mean-field transition; therefore, the pure liquid crystal order parameter data were fitted to the Landau-Ginzburg mean-field model

$$\Phi = \phi_0 \left[ \left(1 + \frac{T_{AC} - T}{T_{AC} - T_{CO}}\right)^{1/2} - 1 \right]^{1/2} \qquad (3)$$

where $T_{CO}$ is the crossover temperature from tricritical to mean-field behavior as $T_{AC}$ is approached. [33] The pure $\overline{8}S5$ data were well described by the Landau-Ginzburg model and gave a SmA-SmC transition temperature $T_{AC}^o = 329.17 \pm 0.02$ K for the pure LC and a crossover temperature $T_{CO} = 327.98 \pm 0.60$ K as adjustable fit parameters. Fits to the above Landau-Ginzburg mean-field form converged for the samples $\rho_S$ = 0.025, 0.039 and 0.060. In the higher density LC + aerosil gels, $\rho_S$ = 0.092 and 0.264, where there is a linear variation over a broad coexistence region as shown in Fig. 5 (e) and (f), the tilt-angle data could not be fit with the mean-field model. Similar order parameter behavior was observed by Cordoyiannis *et. al.* [9] for stiff CE8 + aerosil gels with $\rho_S$ = ~0.18.

It should be stressed that for $\rho_S$ = 0.025, 0.039 and 0.060 samples, a pair of separate and distinct coexisting SmA and SmC peaks could be clearly distinguished in the coexistence region. However, for the higher density gels with $\rho_S$ = 0.092 and 0.264, which correspond to "stiff" gels with concomitantly higher disorder, multiple overlapping SmC peaks were observed in the coexistence region, as described above. [31] For temperatures below the SmA + SmC coexistence region, the position of a single SmC peak continuously evolves toward higher $q$-vectors ( and thus larger tilt angles) as shown in Fig. 5.

The coexistence region presented here differs from a classical two-phase coexistence region in that an exchange of the scattering intensities of the two phases does not obtain. An important implication of this behavior is that the SmA + SmC coexistence region does not arise from an underlying first-order transition but rather is a direct result of the quenched random field disorder. We propose that there is a random distribution of pores with different local values of $T_{AC}(i)$ depending on the size and character of pore $i$. Thus at any given temperature lying in the coexistence region, some pores contain Sm*C* LC while others contain LC molecules that are still in the Sm*A* phase. The behavior of the multipeak Sm*C* structure in the coexistence region reflects this distribution with varying $T - T_{AC}(i)$ values and thus varying intensities $I(i)$ and tilt angles $\Phi(i)$. Below the lower bound of the coexistence temperature range, the Sm*A* scattering peak is absent and the scattering intensity due to Sm*C* ordering from all of the domains merges into a single Sm*C* feature for low density gels, as shown in Fig. 4(d). Such a domain distribution effect would be especially significant for the stiff gels formed by high $\rho_S$ samples and would account for the very broad coexistence range for such samples and the broad asymmetric multi-peak scattering observed at low temperatures. The fact that this Sm*C* scattering is a set of overlapping peaks rather than a very broad but featureless single peak suggests that the distribution of domains is limited to relatively few types rather than a continuum. Naturally, there is no way to decide whether the Sm*A* to Sm*C* conversion in a given "pore" is described by the Landau-Ginzburg model.

It should be noted that high-resolution calorimetric data on $\overline{8}S5$+sil samples [16] show that the *N*-Sm*A* transitions exhibit a critical variation in the excess heat capacity $C_p$ that, except for a narrow region of finite-size rounding, is as well-defined as that for the pure LC. Furthermore, the excess heat capacity associated with the Sm*A*-Sm*C* transition for LC+sil samples is smeared and rounded relative to the Landau mean-field behavior seen in pure $\overline{8}S5$, in qualitative agreement with the X-ray behavior observed here. Thus there is a dramatic difference in the effect of the gel on Sm*A* and Sm*C* ordering. The silica gel creates a quenched random field, but the strength of this field is weak for Sm*A* ordering at the *N*-Sm*A* transition but strong for Sm*C* ordering at the Sm*A*-Sm*C* transition. This seems a reasonable consequence of the pinning of LC molecules at the surface of the sil particles. Such pinning determines the average direction of the nematic director in any given pore. When the Sm*A* phase forms, the normal to the smectic layers should lie parallel to the direction of the director in each pore in order to minimize elastic strain. When SmC forms in the pore, the layer normal must tilt, which requires the concerted sliding of many LC molecules whose long axes will still lie preferentially parallel to the director direction established for that pore. Such a reorganization will be retarded if the aerosil pinning is strong enough, as it would be for philic aerosils. The effect would be especially large for stiff gels with high $\rho_S$ values and small pores.

Clearly the nature of the SmC ordering observed in the present study differs from that of Clegg *et al*. [8] who reported sharp second order SmA-SmC phase transitions in $\overline{8}S5$ + aerosil gels identical to that in the pure material. The present work seems to resolve the mystery of why the $\overline{8}S5$ + aerosil gel behavior reported in Ref. 8 differs so drastically from that observed in other smectic LC-aerosil gel systems. Note that those earlier X-ray results are also inconsistent with the calorimetry measurements of Roshi *et. al*. [16] Specifically, the previous X-ray work indicated that the QRD had little observable effect on the SmA and SmC phases and phase transitions in contrast to the results we have obtained here. The simplest, albeit speculative, explanation of these differing results is that in the Clegg *et. al*. experiments the $\overline{8}S5$ + aerosil materials studied were not fully mixed so that there was some residual pure $\overline{8}S5$ in the samples studied. Because of the sharpness of the pure system X-ray peaks, the pure $\overline{8}S5$ signal would totally dominate the measured profiles, hence leading to erroneous results. In addition, imperfect mixing would also explain the unusually large SmA $\xi_\parallel$ values deep in the SmA phase that were reported in Ref. 8.

**IV. SUMMARY**

In summary, high-resolution X-ray scattering has been performed on a series of $\overline{8}S5$ thermotropic liquid crystals embedded in aerosil gels which covered a broad range of aerosil densities and, hence, quenched random field disorder. The Sm*A* correlation length was shown to exhibit behavior quantitatively consistent with that observed previously in other LC + aerosil systems. The data suggest that the limiting Sm*A* correlation length varies inversely as the aerosil density. Quenched random disorder introduced by the aerosil gels has a drastic effect on the Sm*A* to Sm*C* phase transition and the tilt angle behavior observed. There is a substantial Sm*A* + Sm*C* coexistence region as the system converts from all Sm*A* to all Sm*C*, and for the stiff gels obtained with large $\rho_S$ values this region is very wide and unusual in character.

## ACKNOWLEDGMENTS

We would like to thank Bart Johnson, Martin George and Sean Brennan of the SSRL for technical assistance and Germano Iannacchione for discussions regarding the Sm*A*-Sm*C* transition. This work was supported by the Office of Basic Energy Sciences, U.S. Department of Energy, under Contract No. DE-ACO2-05CH11231. Portions of this research were carried out at the Stanford Synchrotron Radiation Lightsource, a Directorate of SLAC National Accelerator Laboratory and an Office of Science User Facility operated for the U.S. Department of Energy Office of Science by Stanford University. Funding was also provided by the University of California, Berkeley.

**B. Freelon *et. al.,* Figure List**

FIG. 1. Schematic diagrams of the Sm$A$ and Sm$C$ liquid crystal phases (a). Also given are the phase sequence (b), molecular shape (c), and chemical structure (d) of the liquid crystal 4-*n*-pentylphenylthiol-4'-*n*-octyloxybenzoate ($\overline{8}S5$).

FIG. 2. (Color online) Scattering intensity *I(q)* collected from $\overline{8}S5$ + aerosil gels deep in the Sm$A$ phase. The sample temperature was equal to 330.65 K. Solid lines show the total fit of the scattering structure factor, plus the background. In each panel, $S^{thermal}$ is given by the dashed (red) lines while the $S^{static}$ contribution is represented by dotted (blue) lines. The FWHM of the profiles is observed to broaden with increasing aerosil density which indicates that the aerosil diminishes the correlation length of the Sm$A$ order.

FIG. 3 Universal behavior of the Sm$A$ correlation length deep in the Sm$A$ phase versus the aerosil mass density for $\overline{8}S5$ + aerosil. For $\overline{8}S5$ and 4O.8 [18] the quantity plotted $\tilde{\xi}_{\|2}$ is taken to be independent of temperature. For 8CB [3] and 8OCB [7*b*] the plotted quantity is the low-temperature limit of $\xi_{\|}$, where it was assumed that $\xi_{\|2} = \xi_{\|}$ for fits in the Sm$A$ phase.

FIG. 4. Longitudinal $q$-scans taken of the (a) Sm$A$ peak at 330.65 K, (b) Sm$A$ + Sm$C$ coexistence peaks at 325.5 K, (c) Sm$A$ + Sm$C$ coexistence peaks at 325 K, and (d) a Sm$C$ peak at 318 K, in $\overline{8}S5$ + aerosil $\rho_S = 0.025$ g/cm$^3$. The intensity is plotted versus the scattering angle $2\theta$, where $q = (2\pi/\lambda)\sin 2\theta$.

FIG 5. Sm$C$ order parameter (tilt angle) versus temperature for various aerosil mass densities $\rho_S$. The vertical lines enclose Sm$A$ + Sm$C$ coexistence regions. For the pure $\overline{8}S5$, the arrow indicates the value of $T_{AC}^o$ from a fit with Eq. (3).

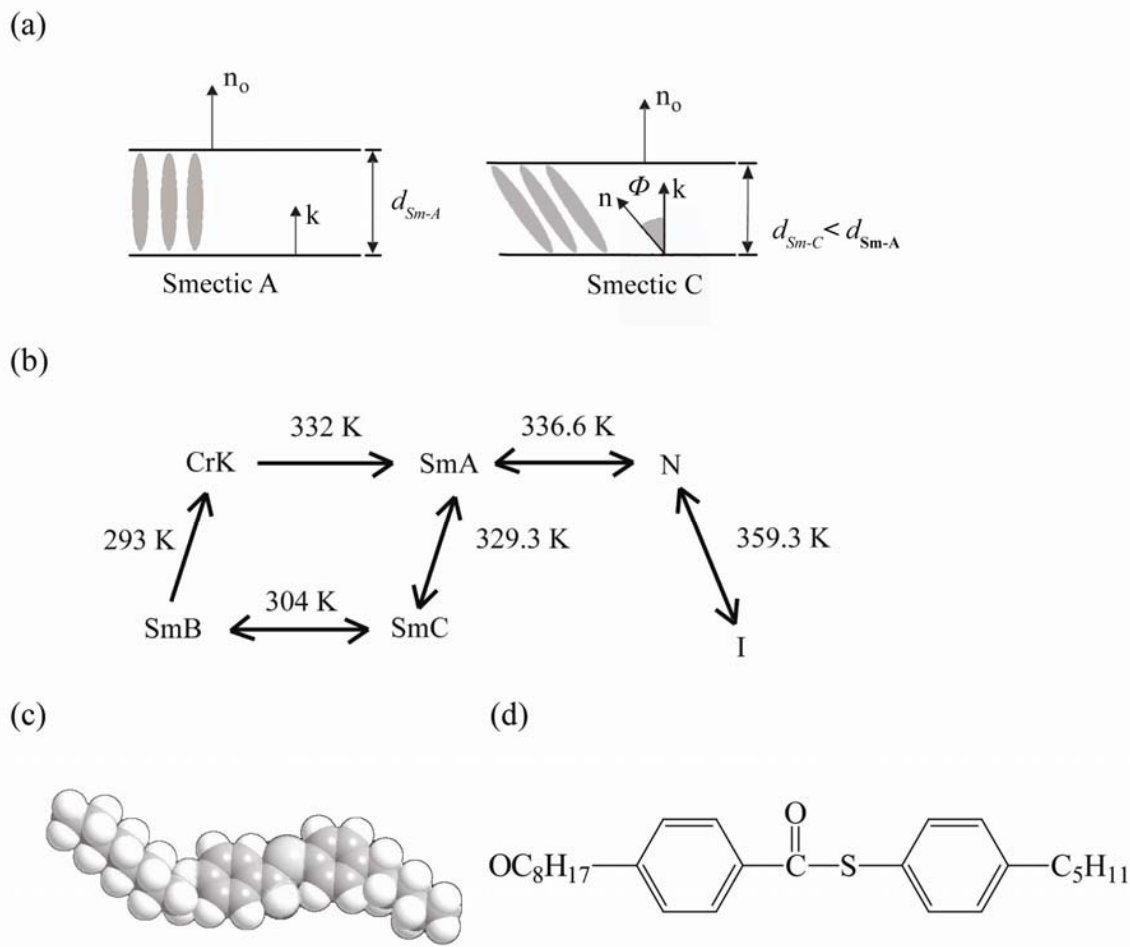

FIG. 1. Schematic diagrams of the Sm*A* and Sm*C* liquid crystal phases (a). Also given are the phase sequence (b), molecular shape (c), and chemical structure (d) of the liquid crystal 4-*n*-pentylphenylthiol-4'-*n*-octyloxybenzoate ($\bar{8}S5$).

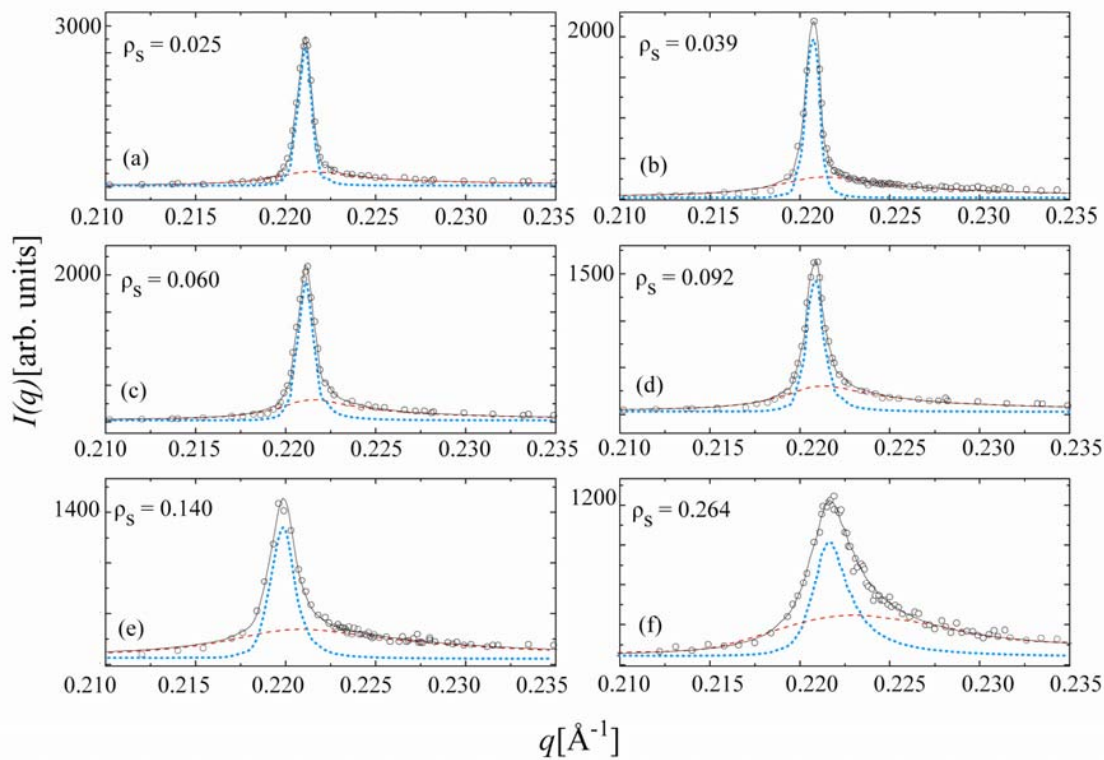

FIG. 2. (Color online) Scattering intensity $I(q)$ collected from $\overline{8}S5$ + aerosil gels deep in the Sm$A$ phase. The sample temperature was equal to 330.65 K. Solid lines show the total fit of the scattering structure factor, plus the background. In each panel, $S^{thermal}$ is given by the dashed (red) lines while the $S^{static}$ contribution is represented by dotted (blue) lines. The FWHM of the profiles is observed to broaden with increasing aerosil density which indicates that the aerosil diminishes the correlation length of the Sm$A$ order.

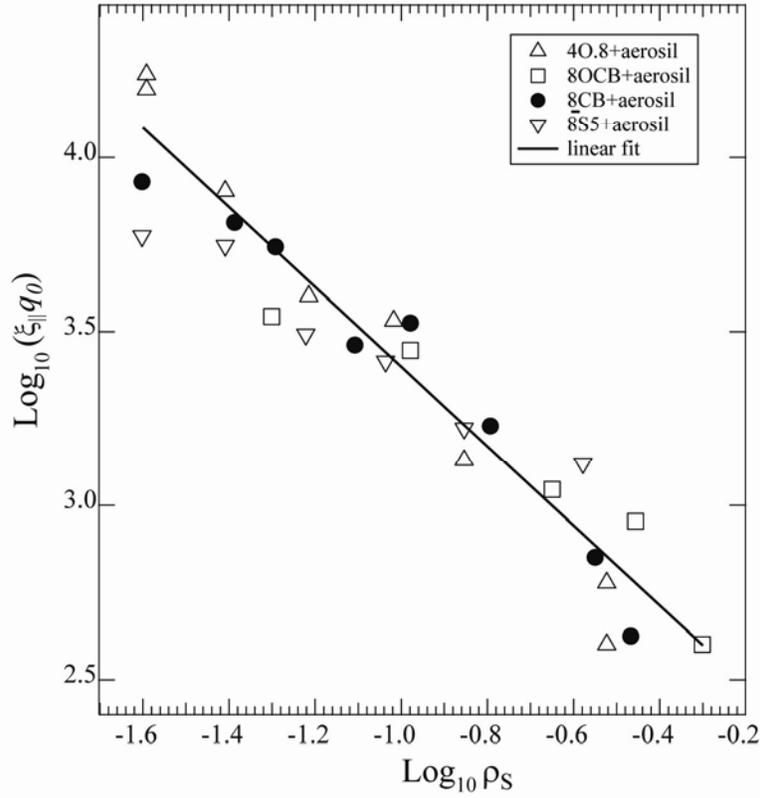

FIG. 3 Universal behavior of the Sm*A* correlation length deep in the Sm*A* phase versus the aerosil mass density for $\overline{8}S5$ + aerosil. For $\overline{8}S5$ and 4O.8 [18] the quantity plotted $\tilde{\xi}_{\|2}$ is taken to be independent of temperature. For 8CB [3] and 8OCB [7b] the plotted quantity is the low-temperature limit of $\xi_\|$, where it was assumed that $\xi_{\|2} = \xi_\|$ for fits in the Sm*A* phase.

**Freelon et. al., FIG. 4**

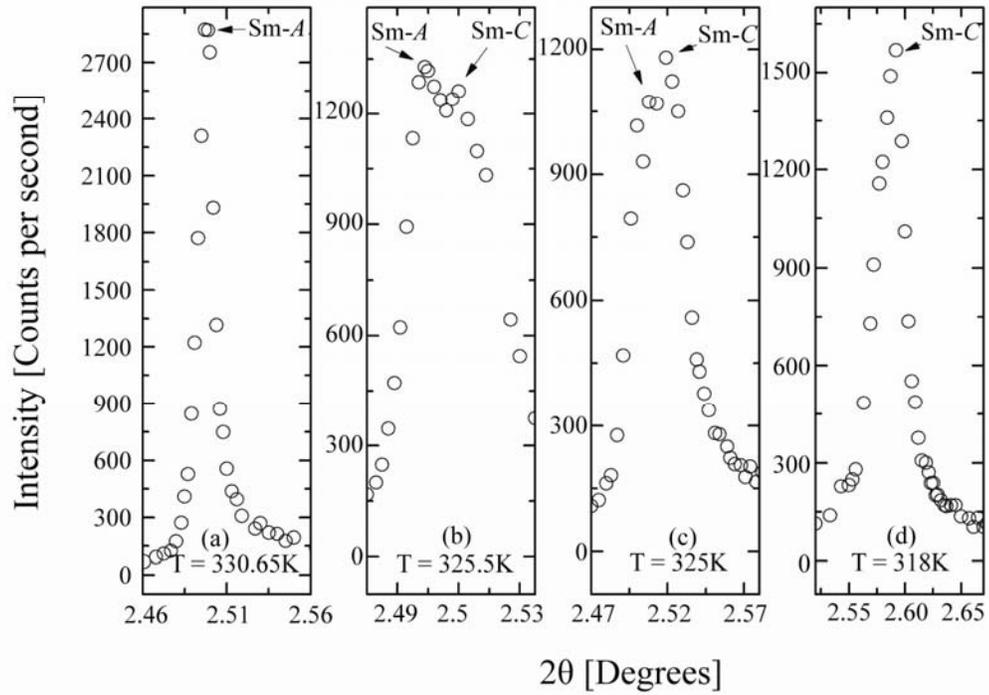

FIG. 4. Longitudinal $q$-scans taken of the (a) Sm$A$ peak at 330.65 K, (b) Sm$A$ + Sm$C$ coexistence peaks at 325.5 K, (c) Sm$A$ + Sm$C$ coexistence peaks at 325 K, and (d) a Sm$C$ peak at 318 K, in $\overline{8}S5$ + aerosil $\rho_S$ = 0.025 g/cm$^3$. The intensity is plotted versus the scattering angle $2\theta$, where $q = (2\pi/\lambda)\sin 2\theta$.

Freelon *et. al.*, FIG. 5

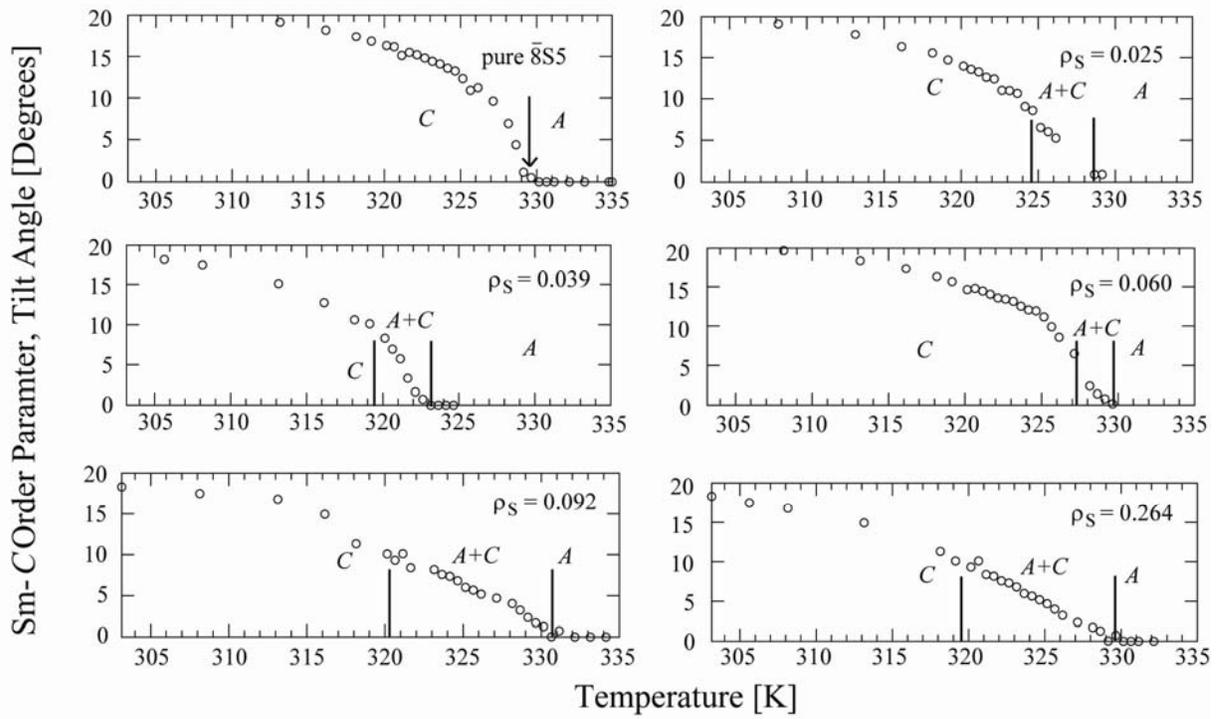

FIG 5. Sm*C* order parameter (tilt angle) versus temperature for various aerosil mass densities $\rho_S$. The vertical lines enclose Sm*A* + Sm*C* coexistence regions. For the pure $\bar{8}S5$, the arrow indicates the value of $T_{AC}^o$ from a fit with Eq. (3).